\documentclass[twoside,twocolumn,tightenlines, showpacs, aps, pre]{revtex4}
          
\usepackage{graphicx}
\usepackage{amssymb}
\usepackage{amsmath}
\usepackage{mathrsfs}

\usepackage{subfigure,wrapfig}
\usepackage{color}
\newcommand{\be}{\begin{equation}}
\newcommand{\ee}{\end{equation}}

\usepackage{ulem}

\begin{document}

\title{Universality Classes of Fluctuation Dynamics in Hierarchical Complex Systems}

\author{A. M. S. Mac\^edo$^{1}$, Iv\'an R. Roa Gonz\'alez$^{1}$, D. S. P. Salazar$^2$ and G. L. Vasconcelos$^1$}
\affiliation{$^1$Departamento de F\'{\i}sica, Laborat\'orio de F\'{\i}sica Te\'orica e Computacional,Universidade Federal de Pernambuco 50670-901 Recife, Pernambuco, Brazil}
\affiliation{$^2$Unidade de Educa\c{c}\~ao a Dist\^ancia e Tecnologia, Universidade Federal Rural de Pernambuco, 52171-900
Recife, PE, Brazil}
\date{September 2016}

\begin{abstract}
A unified approach is proposed to describe the statistics of the short time dynamics of multiscale complex systems. The probability density function of the relevant time series (signal) is represented as a statistical superposition of a large time-scale distribution weighted by the distribution of certain  internal  variables 
that characterize the slowly changing background. The dynamics of the  background  is  formulated as a hierarchical stochastic model whose form is derived from simple physical constraints, which in turn restrict the dynamics to only two possible classes. The probability distributions of both the signal and the background have simple representations in terms of Meijer $G$-functions. The two universality classes for the background dynamics manifest themselves in the signal  distribution as two types of tails: power law and stretched exponential, respectively. A detailed analysis of empirical data from classical turbulence and financial markets shows excellent agreement with the theory.
\end{abstract}

\maketitle

\section{Introduction}

Complex phenomena are known to exhibit stationary time series which often show large deviations from Gaussian statistics. A common procedure to produce non-Gaussian tails is to allow for the violation of one of the assumptions of the central limit theorem, usually the condition that the increments have finite variance which leads to stable L\'evy distributions \cite{levy}. The physically undesirable consequence that  such  distributions have infinite variance is usually dealt with, {\it a posteriori}, by imposing some sort of truncation, generating the so-called truncated L\'evy distributions \cite{truncatedlevy}. In other cases, heavy-tailed distributions can be accounted for by a superposition of two statistics---a procedure known in mathematics as compounding \cite{compounding} and in physics as superstatistics \cite{beck}. Notwithstanding their  empirical successes in several areas \cite{appliedlevy,andrews_1989, castaing_PhysD90,chabaud_PRL94,beck}, such ad hoc treatments of non-Gaussian effects are not completely satisfactory in that they offer no clear physical explanation for the underlying  dynamics driving the fluctuation phenomena.  Here we propose a general stochastic dynamical framework whereby heavy-tailed distributions  naturally emerge. 
It is shown in particular that only two classes of  distributions are allowed as characterized by the nature of the tails: power law and stretched exponential, respectively.
\par
 In this work we also address another  important problem that one often faces when dealing with  fluctuation phenomena in  complex systems, namely the fact the empirical data   can almost equally well be fitted 
 with different  probability distributions, making it difficult to select between  competing models \cite{beck-cohen-swinney,sornette_review}.
 We tackle this model selection problem by introducing a joint fitting procedure whereby we simultaneously fit the empirical distributions of {\it both} the  measured quantity (signal) and an internal variable (background), which represents the changing local environment.
In our formalism,  the background  is  described by a hierarchical stochastic model whose form is severely constrained by basic physical requirements, yielding only two `universality classes' of possible dynamics. Analytical expressions are obtained for the distributions of both the signal and the background  in terms of certain special transcendental functions---the Meijer $G$-functions. The availability of closed form solutions within the same family of special functions not only offers a unified theoretical treatment to the problem,
 but more importantly 
 it allows one to implement  the joint fitting procedure described above, which in turn helps to discriminate between the allowed distributions and also provides a direct check on the  assumptions of the model. Excellent agreement with the theory is found in a detailed analysis of  turbulence data  as well as of financial asset price fluctuations.	Owing to its generality, our theory could have a wide range of applications across different areas and disciplines.
 
\section{The Multiscale Approach}

Consider a multiscale complex system in a stationary state  where the  probability distribution of the relevant observable,  denoted by  $x$, depends on the spatial or temporal scale with  which the system is observed. The large-scale distribution, i.e., measured at a time scale $\tau_0$ above which fluctuations in $x$ are essentially uncorrelated, is assumed to be known and given by
$P(x|\varepsilon_0)$, where $\varepsilon_0$ is a parameter that characterizes the large-scale stationary 
state (global equilibrium) of the system. To be specific,  we shall assume that at  large scales our system follows a Gaussian distribution with variance $\epsilon_0$. (A Gibbsian distribution with mean $\epsilon_0$ could also be chosen   with similar results.) We  suppose furthermore that the observed quantity $x$ has a much faster dynamics than that of its local environment. This means that over short time periods (during which the environment does not change appreciably) the system follows the same distribution $P(x|\varepsilon)$, but where  $\varepsilon$ now characterizes the `local equilibrium'. Under these assumptions,  we can  write 
\be 
P(x |\varepsilon)=\frac{1}{\sqrt{2 \pi \varepsilon}}\exp \left( -\frac{x^2
}{2 \varepsilon}\right).
\label{eq:Gauss}
\ee 
Here we assume that  $\varepsilon$  is  a fluctuating quantity but one that varies  more slowly (in time and space) than $x$. Physically, the quantity  $\varepsilon$ can be identified, for example,  with the local energy flux 
in a  turbulent flow \cite{SV1} or as a `local temperature' in a hierarchical system in thermal equilibrium \cite{SV2}.  
If we  sample the system at  short intervals 
but over a long time span (comparable to
 $\tau_0$), then the statistics of $x$ will be described by the marginal distribution 
 \be
 P(x)=\int_0^\infty P(x|\varepsilon)f(\varepsilon)d\varepsilon,
 \label{eq:Px}
 \ee  
 where $f(\varepsilon)$ is the stationary probability density function of the background variable $\varepsilon$.
 
 One distinctive aspect of our approach  is that we wish to determine the possible  distributions for $f(\varepsilon)$ from general physical arguments, rather than  prescribe  it {\it a priori} as is normally done in the compound and superstatistics formalisms.  Another important ingredient  
is that we seek a theory that  incorporates  multiple time scales---a common feature of  complex systems---and   that  clearly exhibits the connection between the local equilibrium variable $\varepsilon$ and its large scale counterpart $\varepsilon_0$. To this end, the Salazar-Vasconcelos model \cite{SV1,SV2} recently introduced to describe intermittency in multiscale fluctuation phenonema is a natural starting point. 
Our approach is also akin in spirit to the Palmer-Stein-Abrahams-Anderson (PSAA) hierarchical model   for  relaxation in spin glass \cite{anderson}. 

\section{The Dynamical Model}

Let us assume that the system has $N$ well-separated time scales $\tau_i$, $i=1,...,N$, in addition to the 
large scale $\tau_0$,
and let us order them from smallest to largest, i.e., $\tau_i\ll \tau_{i-1}$.  Our arguments will be presented in a general setting, but to fix the ideas one may think of the system as a particle diffusing in a slowly changing environment. The relevant observable $x$ (say, the particle velocity) is measured at the shortest  time scale $\tau_N$, while the environment seen by the particle is described by a slower
 degree of freedom (related to the local diffusion coefficient) represented by a stochastic variable $\varepsilon_N$ with distribution $f_N(\varepsilon_N)$. 
In order to find  $f_N(\varepsilon_N)$, we use a  hierarchical  dynamical model where at each level $i$ $(i=1,...,N$) of the hierarchy the corresponding  variable $\varepsilon_i$ is described by the following  stochastic differential equation  (SDE):
\be 
{d\varepsilon _i}=F( \varepsilon _{i},\varepsilon _{i-1}) dt+G(
\varepsilon_{i},\varepsilon_{i-1})dW_i, 
\label{eq:de}
\ee 
where $F( \varepsilon _{i},\varepsilon _{i-1})$ represents the deterministic driving term, $G(
\varepsilon_{i},\varepsilon_{i-1})$ is the noise amplitude, and the $dW_i$'s  are independent Wiener processes. That the functions $F$ and $G$ depend only on $\varepsilon_{i}$ and $\varepsilon_{i-1}$ encodes the hierarchical nature (local interactions) of the system. It expresses the fact  that the physical constraints imposed on the system  at the large scale are not directly felt at the small scales
but  rather are transferred down the hierarchy through the intervening  scales. 

The possible functional forms of  $F( \varepsilon _{i},\varepsilon _{i-1})$ and $G(\varepsilon_{i},\varepsilon_{i-1})$ are severely constrained by three general physical requirements: (i) {\it equilibrium condition}, which states that 
$\left\langle  \varepsilon _i(t)\right\rangle =\varepsilon_0$ for $t\to\infty$; (ii) {\it invariance under change of scale},
which requires that $F_i(\lambda \varepsilon _{i-1},\lambda \varepsilon _i) =\lambda
F_i(\varepsilon _{i-1},\varepsilon _i) $ and $G_i(\lambda \varepsilon _{i-1},\lambda \varepsilon _i) =\lambda
G_i(\varepsilon _{i-1},\varepsilon _i)$; and (iii) {\it positivity} of  $\varepsilon_i$, meaning that $
{\rm Prob}(\varepsilon _i(t)<0)=0,\;\forall t,\;{\rm if}\;\varepsilon _j(t=0)\geq 0, \forall j$, which in turn entails that  $G_i(0,0) =0$. 
 
The most general model  that satisfies the three requirements above is of the form
\be 
{d\varepsilon _i}=-\gamma _i\left( \varepsilon _{i}-\varepsilon _{i-1}\right) dt+\kappa _i
\varepsilon_{i}^s\varepsilon_{i-1}^{1-s}dW_i, \quad i=1,...,N,
\label{eq:genSDE}
\ee 
where $\gamma_i$ and $\kappa_i$ are positive constants, and $s$ is,  in principle, an arbitrary real number $0\le s\le1$. The positivity of $\varepsilon_i(t)$ is proved in Appendix A. The stationary  distribution $f(\varepsilon_i|\varepsilon_{i-1})$  can be computed  from the  Fokker-Planck equation associated with the SDE (\ref{eq:genSDE}), for  $\varepsilon_{i-1}$ kept fixed. If we now  impose the (physically reasonable) condition that the corresponding Fokker-Planck equation should have analytic coefficients, then this further restricts the model to {\it only two possible  cases}: $s=1$ or $s=1/2$.

In the case $s=1$ (which was first treated in \cite{SV1}), 
the stationary  conditional distribution $f(\varepsilon_i|\varepsilon_{i-1})$ is given by an inverse-gamma distribution
\be
 f(\varepsilon_i|\varepsilon_{i-1}) = \frac{{(\beta_{i} \varepsilon_{i-1})}^{\beta_{i}+1}}{\Gamma (\beta_{i}+1) } {\varepsilon_i^{-\beta_{i}-2}}  e^{{-\beta_{i} \varepsilon_{i-1}}/{\varepsilon_i}}, \label{eq:invgamma}
\ee
whereas for $s=1/2$ the result is a gamma distribution:
\be
 f(\varepsilon_i|\varepsilon_{i-1}) = \frac{{(\beta_{i}/ \varepsilon_{i-1})}^{\beta_{i}}}{\Gamma (\beta_{i}) } {\varepsilon_i^{\beta_{i}-1}}  e^{{-\beta_{i} \varepsilon_{i}}/{\varepsilon_{i-1}}}, \label{eq:gamma}
\ee
where  $\beta_{i}= {2\gamma_i}/{k^{2}_{i}}$ in both cases.

\subsection{Stationary Solutions}

Explicit stationary solutions for $f(\varepsilon_N)$ can be obtained for both models above in the regime of large separation of time scales, i.e., when $ \gamma _N\gg \gamma _{N-1}\gg \ldots \gg \gamma _1$. In this case we can write
\begin{align}
f_N(\varepsilon_N )=\int d\varepsilon _{N-1}\ldots \int d\varepsilon
_1f(\varepsilon _N|\varepsilon _{N-1})\ldots f(\varepsilon _1|\varepsilon _0),
\label{eq:fe}
\end{align}
where $f(\varepsilon _i|\varepsilon _{i-1})$ is  given by either  (\ref{eq:invgamma}) or (\ref{eq:gamma}). A remarkable fact, not noticed before \cite{SV1}, 
is that this  multiple integral can be carried out exactly for both cases, with the result being expressed in terms of the Meijer $G$-function \cite{meijer} (a detailed derivation is shown in Appendix B).  For  the inverse-gamma case ($s=1$) one finds
\begin{equation}
\label{meijer1}
    f_N(\varepsilon_N )= \frac{1}{\varepsilon_0\omega\Gamma(\boldsymbol\beta+{\bf 1})}    G_{ N,0 } ^{ 0,N }  \left( 
\begin{array}{c}
{- \boldsymbol\beta-{\bf 1}}\\ 
-
\end{array}
\bigg |\frac{ \varepsilon_N}{\varepsilon_0 \omega} \right),
\end{equation}
whereas for the gamma case ($s=1/2$) the integral yields
\begin{equation}
\label{meijer2}
    f_N(\varepsilon_N )=
\frac{\omega}{\varepsilon_0\Gamma(\boldsymbol \beta)}    G_{ 0,N } ^{ N,0 }  \left( 
\begin{array}{c}
{-} \\ 
{ \boldsymbol\beta-{\bf 1}}
\end{array}
\bigg |\frac{\omega \varepsilon_N}{\varepsilon_0 }  \right).
\end{equation}
Here $\omega  =\prod_{j=1}^{N}\beta _j$ and we have introduced the vector notation
${\boldsymbol\beta}\equiv (\beta_1,\dots,\beta_N)$ and 
$
\Gamma({\bf a})  \equiv\prod_{j=1}^{N}\Gamma (a _j).
$

It is perhaps worth noting at this point that the  lognormal distribution naturally arises from our multiscale model in the limit  of  infinitely many scales. 
This follows from a direct application of the central limit theorem (CLT) to the variable $\ln \varepsilon_N$, where       $\varepsilon_N=\xi _N\xi _{N-1}\cdots \xi _1$, with
$\xi _i={\varepsilon _i}/{\varepsilon _{i-1}}$.
To apply CLT, we need however to ensure that the second  moment  of $\varepsilon_N$ remains finite for $N\to\infty$,  which is the case for both $s=1$ and $s=1/2$ if we  take the limit  $\beta_i=\beta\to\infty$ in such a way that   $\sigma^2=N/\beta$ stays finite \cite{SV1}. We thus see that the lognormal model often  used in  the compound approach  \cite{castaing_PhysD90,chabaud_PRL94, beck-cohen-swinney} is a limiting case of our multiscale formalism
when infinitely many scales are considered.

\subsection{Marginal Distributions}

The marginal distribution $P_N(x)$ of the signal $x$ measured at the time scale $\tau_N$ can now be computed from the superposition  integral (\ref{eq:Px}), with $P(x|\varepsilon)$ as shown in (\ref{eq:Gauss}). More specifically, we have 
\be 
P_N(x)=\frac{1}{ \sqrt{2\pi }}\int_0^\infty \exp \left( -\frac{x^2%
}{2 \varepsilon_N }\right) {\varepsilon^{-1/2}_N} f_N(\varepsilon_N) d\varepsilon_N,
\label{eq:Psup}
\ee 
where $f_N(\varepsilon_N)$ is  as given    in (\ref{meijer1})  or (\ref{meijer2}). It is again remarkable that this integral can be performed exactly for both  cases, yielding the following  two   classes of
 distributions: 

{ i)  \it Power-law class.} 
This corresponds to the case $s=1$. Upon substituting  (\ref{meijer1}) into  (\ref{eq:Psup})  and using known properties of the $G$-functions \cite{meijer},  the resulting integral can  be expressed also in terms of a $G$-function:
\be 
P_N(x)=
\frac{1}{\sqrt{2\pi\omega\varepsilon_0}\Gamma(\boldsymbol\beta+{\bf 1})}G_{N,1}^{1,N}\left( 
\begin{array}{c}
{ -\boldsymbol\beta-{\bf 1/2}} \\ 
0
\end{array}
\bigg |\frac{x^2}{2\omega \varepsilon_0}\right).
\label{eq:PN1}
\ee 
This formula is an alternative representation of the generalized hypergeometric function ${_N}F_0$ reported in \cite{SV1}.  We note, in particular, that the distributions in this family all have power-law tails \cite{SV1}:
\be 
P_N(x)\sim 
 \sum_{i=1}^{N} \frac{c_{i}}{x^{2\beta_i+3}}, \quad  \mbox{for}\quad |x|\to\infty,
 \label{eq:asym1}
\ee 
 where the $c_i$'s are constants.
 \par

{ii) \it Stretched exponential class.} This class corresponds to  $s=1/2$, in which case the superposition integral (\ref{eq:Psup}) computed with (\ref{meijer2}) yields
\be 
P_N(x)=
\frac{\omega^{1/2}}{\sqrt{2\pi\varepsilon_0}\Gamma(\boldsymbol\beta)}G_{0,N+1}^{N+1,0}\left( 
\begin{array}{c}
- \\ 
{ \boldsymbol\beta-{\bf 1/2}},0
\end{array}
\bigg |\frac{\omega x^2}{2\varepsilon_0}\right) .
\label{eq:PN2}
\ee 
This new class of distributions is a generalization of the $K$-distribution ($N=1$) which is a known  compound distribution with applications to, e.g.,  scattering in random media \cite{radar} and turbulence \cite{andrews_1989,guhr2015}. From the asymptotic expansion of the function $G_{m,n}^{p,q}$, with $m=q=0$ and $n=p=N+1$, one  finds that the tail of the distribution  in this case is given by a modified stretched exponential
\be \
P_N(x)\sim 
{x^{2\theta}}{\exp\left[-(N+1)(\omega x^2/2\varepsilon_0)^{1/(N+1)}\right]},
\label{eq:asym2}
\ee 
 where $\theta=(\sum_{i=1}^N\beta_i -N)/(N+1)$. It is interesting to note that a  stretched exponential also appears (as the long time asymptotics) in the PSAA  model for relaxation in spin glass \cite{anderson}.

 \subsection{The Joint Fitting Procedure.}

 As a first application of our theory we have analyzed two sets of data: i)  velocity  measurements  in a turbulent fluid and ii) 
financial asset  prices. Both systems display hierarchical structures and hence are  natural candidates for applications of our formalism. In turbulence \cite{frisch} one has an energy cascade from large to small length scales, so that our background variables $\epsilon_i$  correspond to the fluctuating energy fluxes between  adjacent scales. In financial markets there is an information cascade from long to short temporal scales \cite{nature1996}, with the background variables representing the stochastic volatilities at the different time scales. Furthermore, the statistics of both velocity and price fluctuations have been modeled by different distributions (often with comparable degrees of agreement, 
see, e.g., \cite{sornette_review,guhr2015,beck_2016}), 
and so they are a good test for our joint fitting procedure (see below) whereby the distributions of both the signal and the background are fitted according to Eqs.~(\ref{eq:PN1}) and (\ref{meijer1}), respectively. (Because the empirical distributions in both cases show a power-law trend it suffices to consider the case $s=1$.) 

\begin{figure}[t!]
\center \subfigure[\label{fig1a}]{\includegraphics[width=0.45\textwidth]{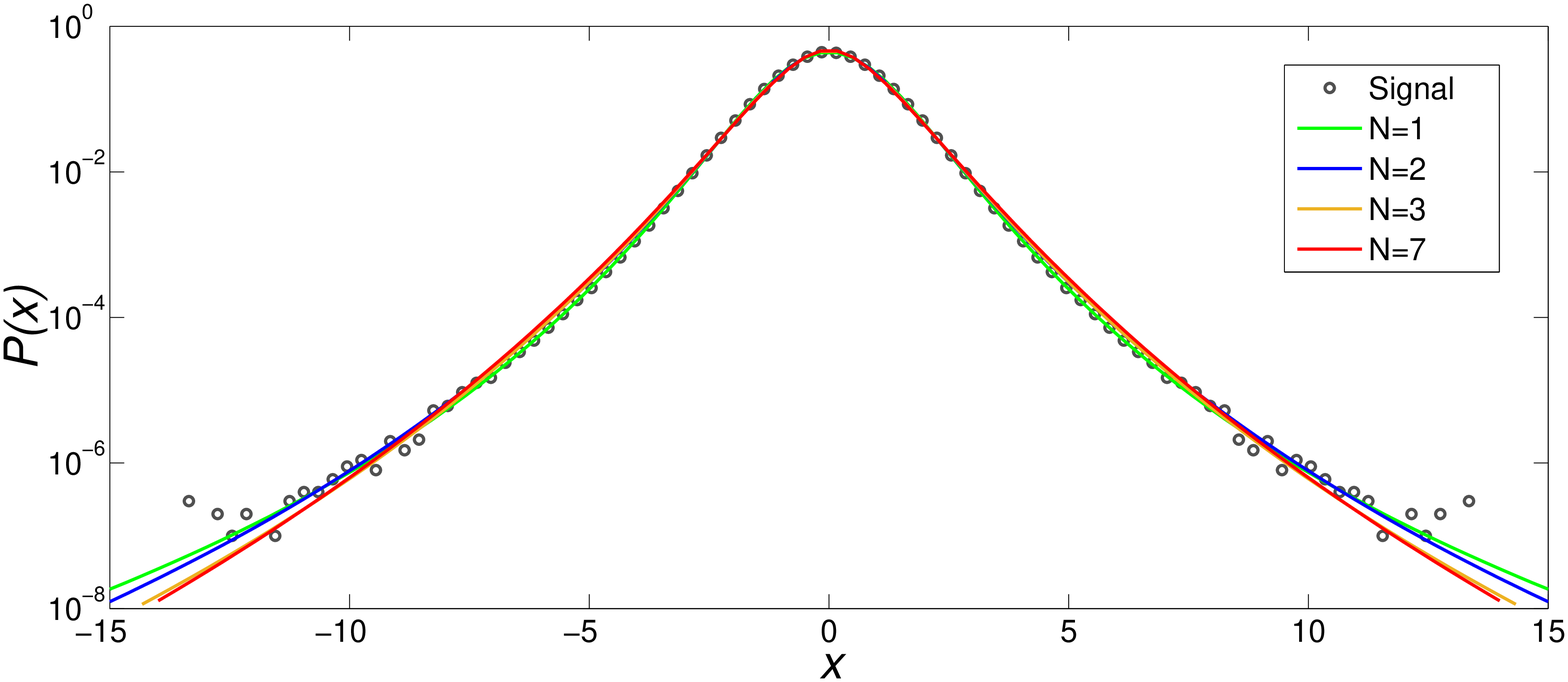}}
\qquad \subfigure[\label{fig1b}]{\includegraphics[width=0.45\textwidth]{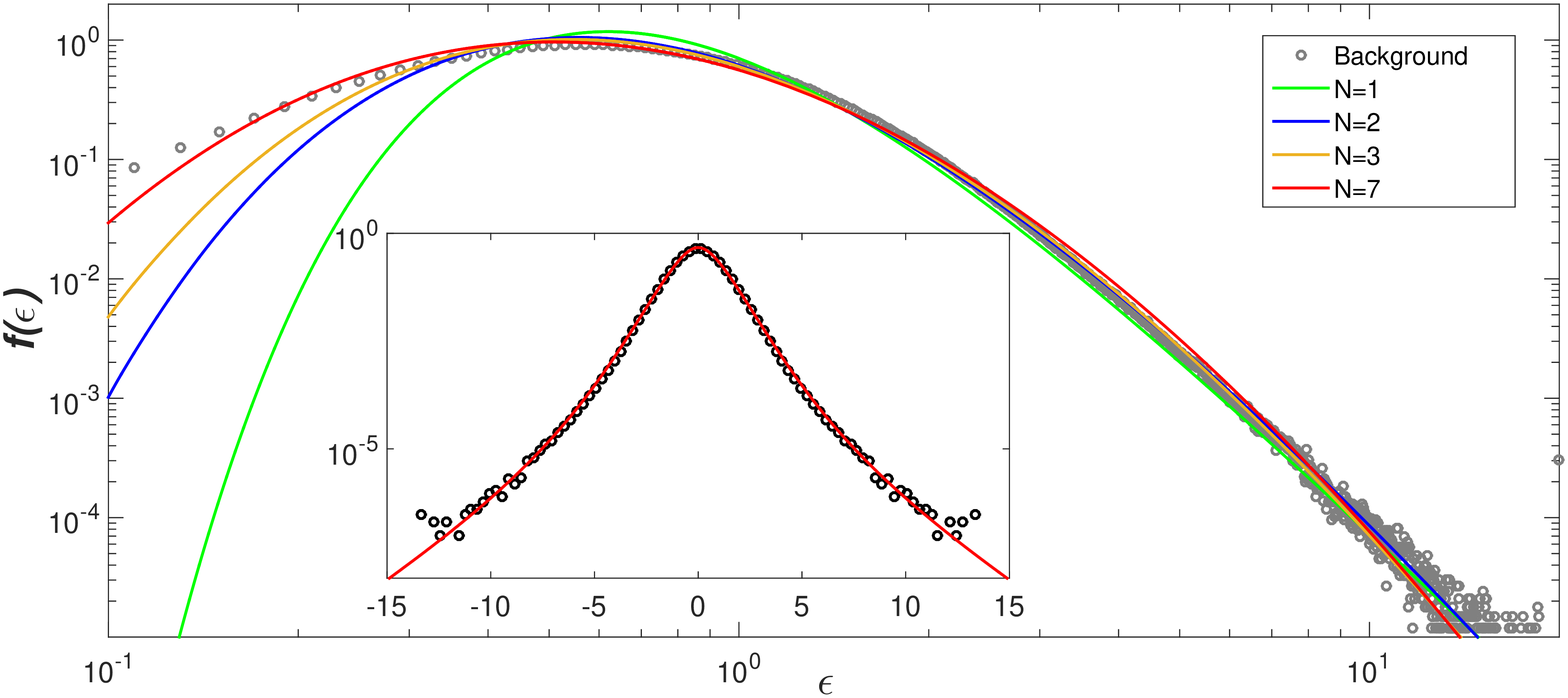}}
\caption{(Color online) (a) Experimental distribution for velocity increments  (black dots) in a turbulent jet flow and model predictions (solid lines) for $N=1$ and $\beta=3.26$ (green), $N=2$ and $\beta=5.16$ (blue), $N=3$ and $\beta=7,47$ (black), $N=7$ and $\beta=15.5$ (red);  (b) histogram (black dots) of the variance series $\epsilon(t)$ and model predictions (solid lines) with same parameters and color conventions as in (a). Inset shows the compounding (red line) of $\epsilon(t)$ with a Gaussian and the experimental distribution (black dots) of velocity increments.}
\end{figure}

In the turbulence data we consider, our  variable $x$ represents  velocity increments in a low temperature gaseous helium jet \cite{french_group}, i.e., $x(t)=u(t+\tau)-u(t)$, where $u$ is the  velocity measured  (by a hot wire probe)  on the axis of the jet and $\tau$ is the inverse of the sampling rate. Here we analyze a data set of $10^{7}$ points obtained from an experimental run performed at  Reynolds number $Re=295000$ and $\tau^{-1}=271.5$ kHz \cite{french_group}. In Fig.~\ref{fig1a} we show the symmetric part of the empirical distribution for the velocity increments together with the best fit by the theoretical formula  $P_N(x)$, given in (\ref{eq:PN1}), for various values of $N$. In performing the fits we set $\beta_i=\beta$ and thus are left with only one parameter to adjust for any given $N$. One sees from Fig.~\ref{fig1a}  that there  appears to be valid solutions for  different values of $N$, and so an unambiguous choice of model distribution cannot be made  on the basis of this comparison alone.

To solve this quandary, we need to extract  the background series directly from the experimental data and examine its distribution.  For that, we first divide our original time series in intervals of size $M$ and for each such interval compute a  variance estimator, 
 $\epsilon(t)=\frac{1}{M}\sum_{j=0}^{M-1}[x(t-j\delta t)-\bar{x}(t)]^2$, where $\bar{x}(t)=\frac{1}{M}\sum_{j=0}^{M-1}x(t-j\delta t)$, thus generating a new time series.   Next, we numerically compound the background series  $\epsilon(t)$ with a Gaussian for various $M$, and select the value of $M$ for which the corresponding superposition integral best fits the  distribution of the original series.  Excellent agreement is  found for $M=19$; see inset of Fig.~\ref{fig1b}. In the main plot of Fig.~\ref{fig1b}  we compare the distribution of $\epsilon(t)$ for this optimal window size with the distribution $f_N(\varepsilon_N)$ given in (\ref{meijer1}) for the same parameters as in Fig.~\ref{fig1a}. Taking into account both  figures \ref{fig1a} and \ref{fig1b}, we  conclude that the solution with $N=7$ and $\beta=15.5$ gives the best overall fit to the turbulence data.

\begin{figure}[t!]
\center \subfigure[\label{fig2a}]{\includegraphics[width=0.45\textwidth]{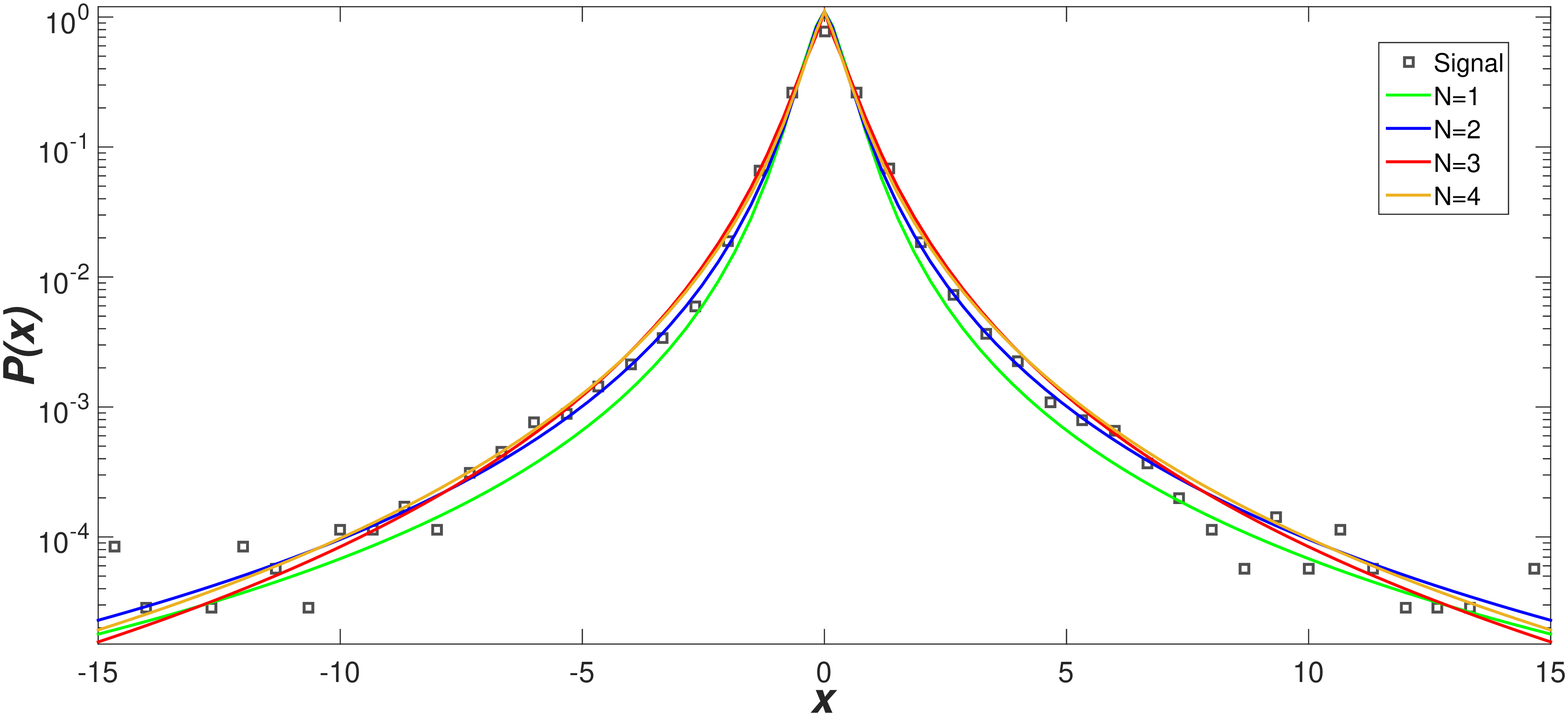}}
\qquad \subfigure[\label{fig2b}]{\includegraphics[width=0.45\textwidth]{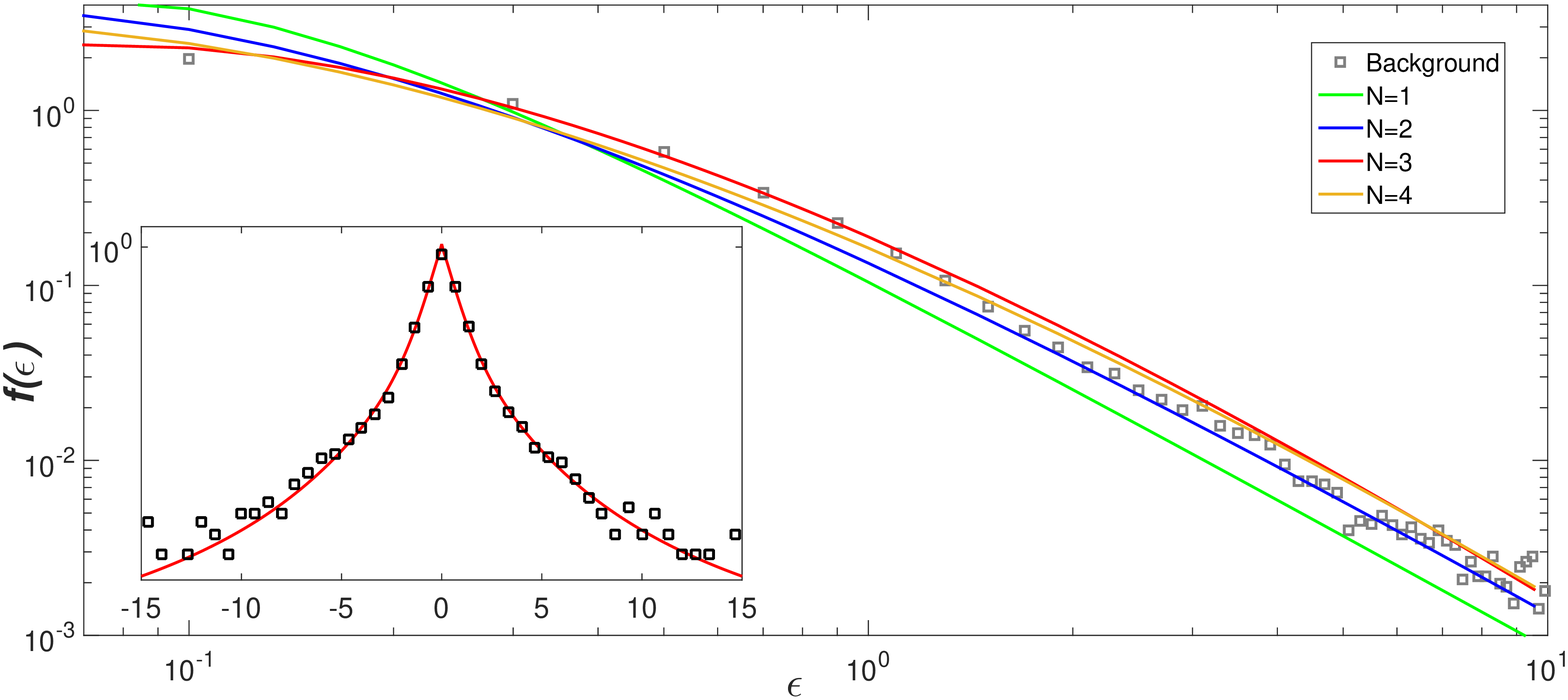}}
\caption{(Color online) (a) Empirical distribution of intraday returns of the Ibovespa index  (black dots) and model predictions (solid lines) for $N=1$ and $\beta=0.15$ (green), $N=2$ and $\beta=0.5$ (blue), $N=3$ and $\beta=1.2$ (black), $N=4$ and $\beta=1.32$ (red);  (b) histogram (black dots) of the variance series $\epsilon(t)$ and model predictions (solid lines) with same parameters and color conventions as in (a). Inset shows the superposition (red line) of $\epsilon(t)$ with a Gaussian and the empirical distribution (black dots).}
\end{figure}

The financial data we analyzed correspond to  the logarithmic intraday returns of the Ibovespa  index of the  S\~ao Paulo Stock Exchange, that is, $x(t)=\ln S(t+\tau)-\ln S(t)$, where $S(t)$ is the  Ibovespa value at time $t$ and   $\tau=30$s. We analyzed quotes from the period of November 2002 to March 2004, corresponding to a total of about 53000 points. The empirical distribution of the returns and respective fits using (\ref{eq:PN1}) are shown in Fig.~\ref{fig2a}. In this case, the optimal window size for the variance series is $M=5$, as indicated in the  inset of Fig.~\ref{fig2b}, with the fits to the variance distribution being shown in the main plot of Fig.~\ref{fig2b}. Using the joint fitting procedure outlined above, we conclude that  the best combined fit occurs for  $N=3$ and $\beta=1.2$. It is interesting to note that for both sets of data considered here we have found $N>1$,  thus  indicating that  fluctuation processes across multiple scales do indeed take place in these systems. (Recall that $N$ is an estimate of the number of relevant scales in the problem.)

\section{Conclusions}

Describing fluctuation phenomena in multiscale complex systems is  an admittedly difficult task because (among other reasons) one does not usually have direct access to the interscale dynamics, and hence indirect inferences have to be made about its effect on the measured quantities. For instance, non-Gaussian statistics is usually seen as an evidence of complex interactions between scales, but a general dynamical framework to explain such deviations from Gaussianity has not yet been established. Here 
we have shown, from a rather minimal set of assumptions on the interscale stochastic dynamics, 
that there exist two general classes of heavy-tailed distributions  for the statistics of multiscale fluctuations. The distributions in both  classes are given in terms of the same family of special functions (Meijer $G$-function) but  differ regarding the nature of the tail: power law and  modified stretched exponential, respectively. Good agreement was found with  experimental data on classical fluid turbulence as well as financial data---both sets of data analyzed here were shown to belong to the power-law class. Further development of the theory presented here and additional applications, including of  the stretched-exponential class,  will be discussed in forthcoming publications.

\begin{acknowledgments}
 We are grateful to  B. Chabaud and P. E. Roche for sharing their turbulence data with us and to  the S\~ao Paulo Stock Exchange for providing the financial data. This work was supported in part by the Brazilian agencies CNPq  and FACEPE.
\end{acknowledgments}

\appendix

\section{Positivity of $\varepsilon_i(t)$}

By hypothesis, the large scale variable $\varepsilon_0$ is positive and $\varepsilon_1(t)$ is a continuous stochastic process. Therefore, if a given realization of (\ref{eq:genSDE}) were to give a negative value of $\varepsilon_1$ at some time $t^{\prime}$, then it would have reached the value zero at an earlier time $t<t^{\prime}$. Setting $\varepsilon_1(t)=0$ in (\ref{eq:genSDE}), we get
\be 
d \varepsilon_1(t)=\gamma_1 \varepsilon_0 >0,
\ee 
which implies that $\varepsilon_1>0$ for the whole process. Iterating this argument, we conclude that if $\varepsilon_{j-1}(t)>0$ then $\varepsilon_j(t)>0$ for all $t$. We close by remarking that the case $\varepsilon_0=0$ yields the trivial fixed point $\varepsilon_j(t)=0$,  $\forall j,t$.

\medskip 

\section{Background Distributions}
We start by introducing the variable
\be 
y=\frac{\varepsilon _N}{\varepsilon _0}=\prod_{j=1}^N\xi _j,
\ee
where $\xi _j=\varepsilon _j/\varepsilon _{j-1}$, then $f_N(\varepsilon_N)=g(y)/\varepsilon_0$ and

\be 
g(y)=\int_0^\infty \prod_{j=1}^Ng_j(\xi _j)d\xi _j\delta (y-\xi _1\cdots \xi
_N),
\label{g}
\ee 
where from (\ref{eq:invgamma}) we get
\be 
g_j(\xi _j)=\frac{\beta _j^{\beta _j+1}}{\Gamma (\beta _j+1)}\xi _j^{-\beta
_j-2}e^{-\beta _j/\xi _j},
\label{eq:invgamma2}
\ee 
whilst from (\ref{eq:gamma}) we obtain
\be 
g_j(\xi _j)=\frac{\beta _j^{\beta _j}}{\Gamma (\beta _j)}\xi _j^{\beta
_j-1}e^{-\beta _j\xi _j}.
\label{eq:gamma2}
\ee 
Now we apply the Mellin transform, defined as

\be 
{\cal M}[g;s] \equiv \int_0^\infty dy y^{s-1}g(y).
\ee 
to both sides of (\ref{g}). We find
\be 
{\cal M}[g;s] =\prod_{j=1}^N{\cal M}[g_j;s],
\label{Mellin_g}
\ee 
where 
\be 
{\cal M}[g_j;s] =\beta _j^{s-1}\frac{\Gamma
(\beta _j-s+2)}{\Gamma (\beta _j+1)}\label{Mellin-invgamma}
\ee 
is the Mellin transform of (\ref{eq:invgamma2}) and
\be 
{\cal M}[g_j;s] =\frac{\Gamma (\beta _j+s-1)}{%
\beta _j^{s-1}\Gamma (\beta _j)}\label{Mellin-gamma}
\ee 
is the Mellin transform of (\ref{eq:gamma2}). Next, we use the following property of the Meijer $G$ function \cite{meijer}. If the Mellin transform of $g(y)$ is
\[
{\cal M}[g;s]=\frac{\alpha ^{-s}\prod_{j=1}^m\Gamma
(s+b_j)\prod_{j=1}^n\Gamma (1-s+a_j)}{\prod_{j=m+1}^q\Gamma
(1-s+b_j)\prod_{j=n+1}^p\Gamma (s+a_j)}
\]
then
\be 
g(y)=G_{p,q}^{m,n}\left( 
\begin{array}{l}
a_1,\ldots ,a_p \\ 
b_1,\ldots ,b_q
\end{array}
\bigg |\alpha y\right) \label{Mellin-G}.
\ee 
Using (\ref{Mellin_g}), (\ref{Mellin-invgamma}), (\ref{Mellin-gamma}) and (\ref{Mellin-G}) we obtain (\ref{meijer1}) and (\ref{meijer2}) respectively.

\medskip
\bibliography{Referencias}

\end{document}